\documentclass[journal]{IEEEtran}

\ifCLASSINFOpdf
\else
   \usepackage[dvips]{graphicx}
\fi
\usepackage{url}

\usepackage{graphicx}

\usepackage{amsmath,amssymb,amsfonts,amsthm,latexsym,mathrsfs,bbm,cite,algorithm,algorithmic}

\DeclareMathOperator{\blkdiag}{blkdiag}
\DeclareMathOperator{\prox}{prox}

\begin{document}

\title{Consistent {ICA}: \\ Determined {BSS} Meets Spectrogram Consistency}

\author{Kohei Yatabe, \IEEEmembership{Member, IEEE}
\thanks{Manuscript received April 22, 2020; revised May 19, 2020; accepted May 19, 2020.}
\thanks{K. Yatabe is with Department of Intermedia Art and Science, Waseda University, Tokyo, Japan (e-mail: k.yatabe@asagi.waseda.jp).}}

\markboth{IEEE Signal Processing Letters, Vol. 27, 2020}
{Yatabe: Consistent ICA}
\maketitle

\begin{abstract}
Multichannel audio blind source separation (BSS) in the determined situation (the number of microphones is equal to that of the sources), or determined BSS, is performed by multichannel linear filtering in the time-frequency domain to handle the convolutive mixing process.
Ordinarily, the filter treats each frequency independently, which causes the well-known \textit{permutation problem}, i.e., the problem of how to align the frequency-wise filters so that each separated component is correctly assigned to the corresponding sources.
In this paper, it is shown that the general property of the time-frequency-domain representation called \textit{spectrogram consistency} can be an assistant for solving the permutation problem.
\vspace{-3pt}
\end{abstract}


\IEEEpeerreviewmaketitle

\section{Introduction}

\IEEEPARstart{B}{lind} source separation (BSS) is the methodology for recovering source signals from multiple mixtures without knowing the mixing system \cite{Comon1994,Smaragdis1998,hyvarinen2000,Buchner2005,Ono2010auxFDICA,Tachikawa2018,BSSreview}.
Let a mixing process of audio signals be approximated in the time-frequency domain:
\begin{equation}
    \mathbf{x}[t,f]\approx \mathbf{A}[f]\mathbf{s}[t,f],
    \label{eq:mixingModel}
\end{equation}
where $\mathbf{x} = [x_1,x_2,\ldots x_M]^{\mathrm{T}}$ is an observation obtained by $M$ microphones, $\mathbf{s} = [s_1,s_2,\ldots s_N]^{\mathrm{T}}$ is the source signals to be recovered, $\mathbf{A}[f]$ is an $M\!\times\!N$ mixing matrix, and $t$ and $f$ are indices of time and frequency, respectively.
Then, the aim of BSS is to recover $N$ source signals $\mathbf{s}$ from the mixtures $\mathbf{x}$.
In a determined or overdetermined situation ($M\!\geq\! N$), the BSS problem is formulated as an estimation problem of finding an $N\times M$ demixing matrix $\mathbf{W}[f]$ which is a left inverse of $\mathbf{A}[f]$ (i.e., $\mathbf{W}[f]\mathbf{A}[f]=\mathbf{I}$).
Then, the source signals are given as
\begin{equation}
    \mathbf{W}[f]\mathbf{x}[t,f]\approx \mathbf{W}[f]\mathbf{A}[f]\mathbf{s}[t,f] = \mathbf{s}[t,f].
    \label{eq:demixing}
\end{equation}
For the sake of simplicity, this paper considers the determined situation ($M\!=\!N$) only.

As the matrix of demixing filter $\mathbf{W}[f]$ in the above model is defined frequency-wise, the frequency-domain BSS suffers from the \textit{permutation problem} \cite{Anemuller2000,Murata2001,Sawada2004Trans}.
Even when each demixing filter admits the perfect separation, the reconstructed result may be mixed up (see Fig.\:\ref{fig:speechPermutationExamples}) because its permutation cannot be determined without the knowledge on the mixing system.
That is, the separated component $\mathbf{W}[f]\mathbf{x}[t,f] \approx \mathbf{P}[f]\mathbf{s}[t,f]$ can be a permuted version of the original source $\mathbf{s}[t,f]$, where $\mathbf{P}[f]$ is an arbitrary frequency-wise permutation matrix.
To reconstruct the source signal, the permutation $\mathbf{P}[f]$ must be the same for all frequency $f$.

To resolve such permutation problem, recent BSS methods explicitly model the inter-frequency relation within the source signal.
For instance, the independent vector analysis (IVA) \cite{Hiroe2006,Kim2007,Ono2011auxIVA} assumes co-occurrence among the frequency components in each source, and the independent low-rank matrix analysis (ILRMA) \cite{Kitamura2015,Kitamura2016,KitamuraEURASIPJ2018} assumes low-rankness on the power spectrogram of each source.
These models are based on the structure of audio signals (e.g., speech and music) in the time-frequency domain and pull information for separation by approximating the magnitude of spectrograms.

In this paper, in contrast to the conventional methods based on the property of source signals, the general property of the time-frequency representation called \textit{spectrogram consistency} \cite{LeRouxGLA,LeRouxWiener,fastGLA,YatabeAST2019} is considered as an assistant for solving the permutation problem.
Roughly speaking, it is an inter-frequency relation closely tied with the smoothness of a spectrogram (see Fig.\:\ref{fig:consistentExamples}) and often utilized in phase-aware signal processing and phase reconstruction \cite{Gerkmann15,Mowlaee16,MowlaeeBook,YatabePC,YatabeReLU,MasuyamaManifold,MasuyamaADMM,MasuyamaDeGLI,MasuyamaHPSS,MasuyamaLowRank,Masuyama2020}.
When the frequency-wise permutation is not well-aligned, the separated signal results in a non-smooth inconsistent spectrogram (see Fig.\:\ref{fig:speechPermutationExamples}), and thus inducing consistency should be helpful for solving the permutation problem.
An example algorithm for doing so is proposed based on the algorithm in \cite{YatabeICASSP2018,YatabeICASSP2019}, and the independent component analysis (ICA) \cite{Comon1994,Smaragdis1998,hyvarinen2000,Buchner2005,Ono2010auxFDICA} is implemented to show the positive effect of the spectrogram consistency.

\section{Consistency and Permutation Problem}
\label{sec:consistencyVSpermutation}

In this section, brief explanation of the spectrogram consistency is presented together with some illustrative examples.

\subsection{Spectrogram Consistency}

Let the short-time Fourier transform (STFT) of a time-domain signal $\xi$ with respect to a window $w$ be defined as%
\footnote{As the spectrogram of $m$th observation is written as $x_m[t,f]$ in Eq.~\eqref{eq:mixingModel}, $\xi_m[l]$ is used for its time-domain counterpart just for now. To make sure that Eq.~\eqref{eq:defSTFT} is the standard inner product, the summation is taken up to the signal length $L$, and thus the length of the window $w$ must also be $L$ but compactly supported within length $F$ (the index of the window $[l\!-\!at]$ is read as $[l\!-\!at]\!\!\!\mod\!L$). See the supporting material of \cite{YatabeAST2019} if any unclear point exists (though the definition of STFT is not important for this paper).}
\begin{equation}
    \mathrm{STFT}_{w}(\xi)[t,f] = \sum_{l=0}^{L-1}\xi[l]\,\overline{w[l-at]\,\mathrm{e}^{2\pi\mathbbm{i}lf/F}},
    \label{eq:defSTFT}
\end{equation}
where the overline indicates complex conjugation, $L$ is the length of the signal to be transformed, and $\mathbbm{i}$ is the imaginary unit.
This linear equation, or the standard inner product, can be written as a matrix-vector product whose matrix is $TF\!\times\!L$ (its explicit form is omitted due to space limitations).
Usually, such matrix associated with STFT is rectangular and thin%
\footnote{Except some special cases, this condition is a requirement for perfectly reconstructing the time-domain signal from its spectrogram, and thus usual.}\,(i.e., $TF\!>\!L$),
and therefore a spectrogram lies on the $L$-dimensional subspace.
In other words, any component in the remaining $(TF\!-\!L)$-dimensional subspace does not related to the time-domain signal because this remaining subspace is the null space of the inverse STFT (denoted by $\mathrm{iSTFT}_{\widetilde{w}}$).

A spectrogram is said to be \textit{consistent} when it does not contain any component in that $(TF\!-\!L)$-dimensional subspace.
To be more specific, let a pair of analysis and synthesis windows ($w$ and $\widetilde{w}$) satisfy the perfect reconstruction condition, i.e.,
\begin{equation}
    \xi = \mathrm{iSTFT}_{\widetilde{w}}(\mathrm{STFT}_{w}(\xi))
\end{equation}
holds for all time-domain signal $\xi$.
While the spectrogram $\mathrm{STFT}_{w}(\xi)$ is always in the aforementioned $L$-dimensional subspace (or the image of STFT denoted by $\mathrm{Im}(\mathrm{STFT}_w)$),
\begin{equation}
    \xi = \mathrm{iSTFT}_{\widetilde{w}}(\mathrm{STFT}_{w}(\xi) + \nu)\qquad(\nu\notin\mathrm{Im}(\mathrm{STFT}_w))
    \label{eq:removeNu}
\end{equation}
also holds because $\mathrm{iSTFT}_{\widetilde{w}}(\nu)=0$ for any $\nu\notin\mathrm{Im}(\mathrm{STFT}_w)$.
A spectrogram is consistent when it is in $\mathrm{Im}(\mathrm{STFT}_w)$ and does not contain any component $\nu$ outside that.
By defining a projection onto the consistent subspace $\mathrm{Im}(\mathrm{STFT}_w)$ as
\begin{equation}
    \mathrm{proj}^{\text{consist}\!}_{w,\widetilde{w}}(x) = \mathrm{STFT}_{w}(\mathrm{iSTFT}_{\widetilde{w}}(x)),
    \label{eq:defProj}
\end{equation}
the consistent spectrogram can be characterized as its fixed point, i.e., $x$ is consistent if and only if $x = \mathrm{proj}^{\text{consist}\!}_{w,\widetilde{w}}(x)$.

\begin{figure}[!t]
    \centering
    \includegraphics[width=0.96\columnwidth]{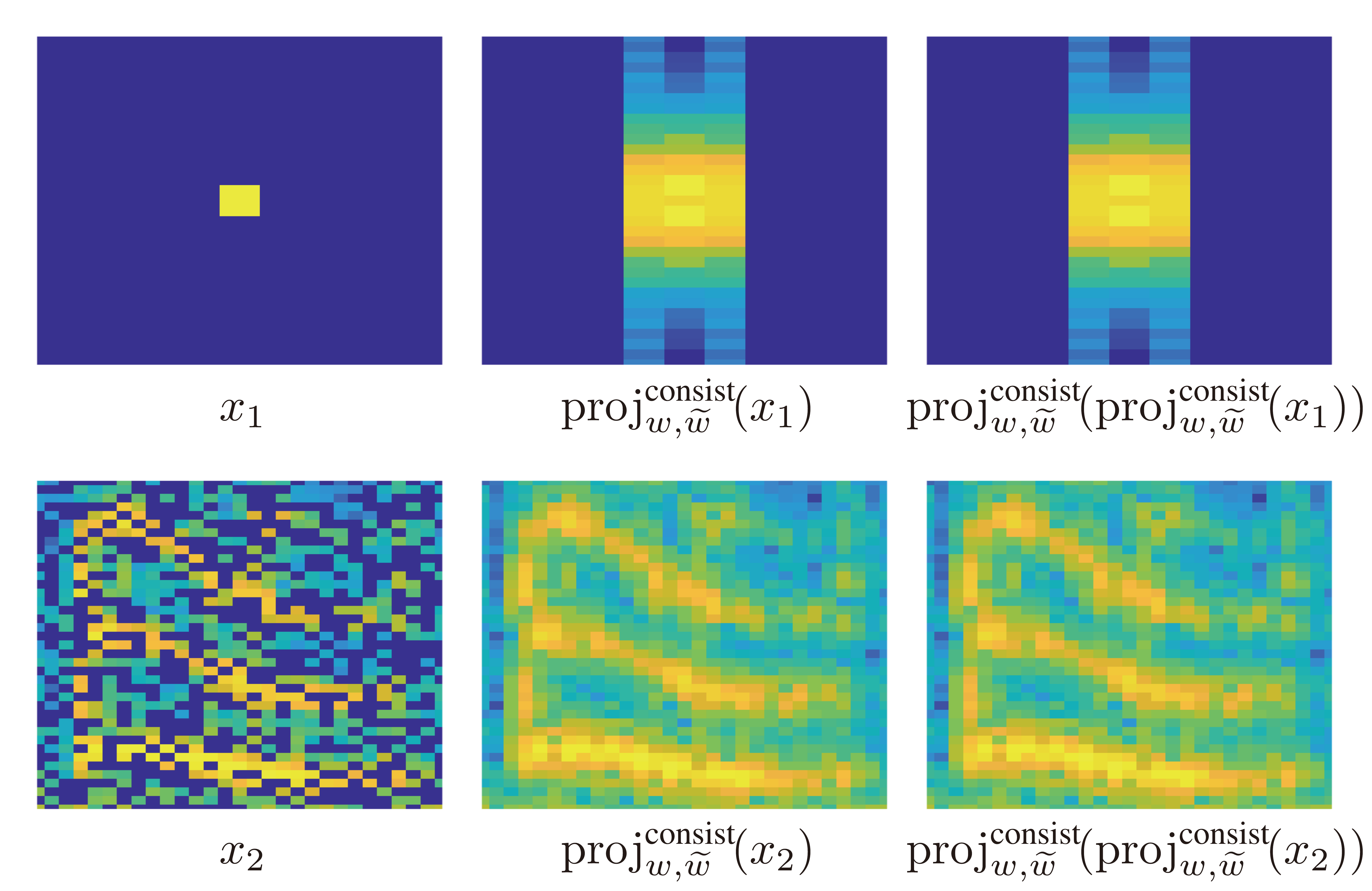}
    \vspace{-6pt}
    \caption{Examples of the consistent/inconsistent spectrograms. The leftmost column shows two synthetic inconsistent spectrograms, while those in the right two columns are consistent owing to the projection in Eq.~\eqref{eq:defProj}.}
    \label{fig:consistentExamples}
\end{figure}

\subsection{Examples of Consistent/Inconsistent Spectrogram}

To demonstrate the spectrogram consistency in an intuitive manner, synthetic inconsistent spectrograms and their consistent counterparts are shown in Fig.\:\ref{fig:consistentExamples} as illustrative examples.
These power spectrograms were calculated with typical parameters (half-overlapping Hann window) and colored by $100$\,dB range.
In the top row, a pulsive spectrogram is shown to illustrate how the projection spreads the component along frequency.
In the bottom row, a spectrogram of a speech signal with random dropout ($50$\,\%) is shown to illustrate how the projection makes a non-smooth spectrogram smooth.
In both cases, the second projection (right column) results in the same spectrogram as that of the first projection (middle column), which confirms the fixed-point characterization.
From these examples, it is clear that enforcing consistency is a smoothing process of a spectrogram in the time-frequency domain.
This smoothing effect is performed by the inverse STFT which removes the cause of non-smoothness ($\nu$ in Eq.~\eqref{eq:removeNu}).

\subsection{Effect of Permutation in Terms of Consistency}

\begin{figure}[!t]
    \centering
    \includegraphics[width=0.96\columnwidth]{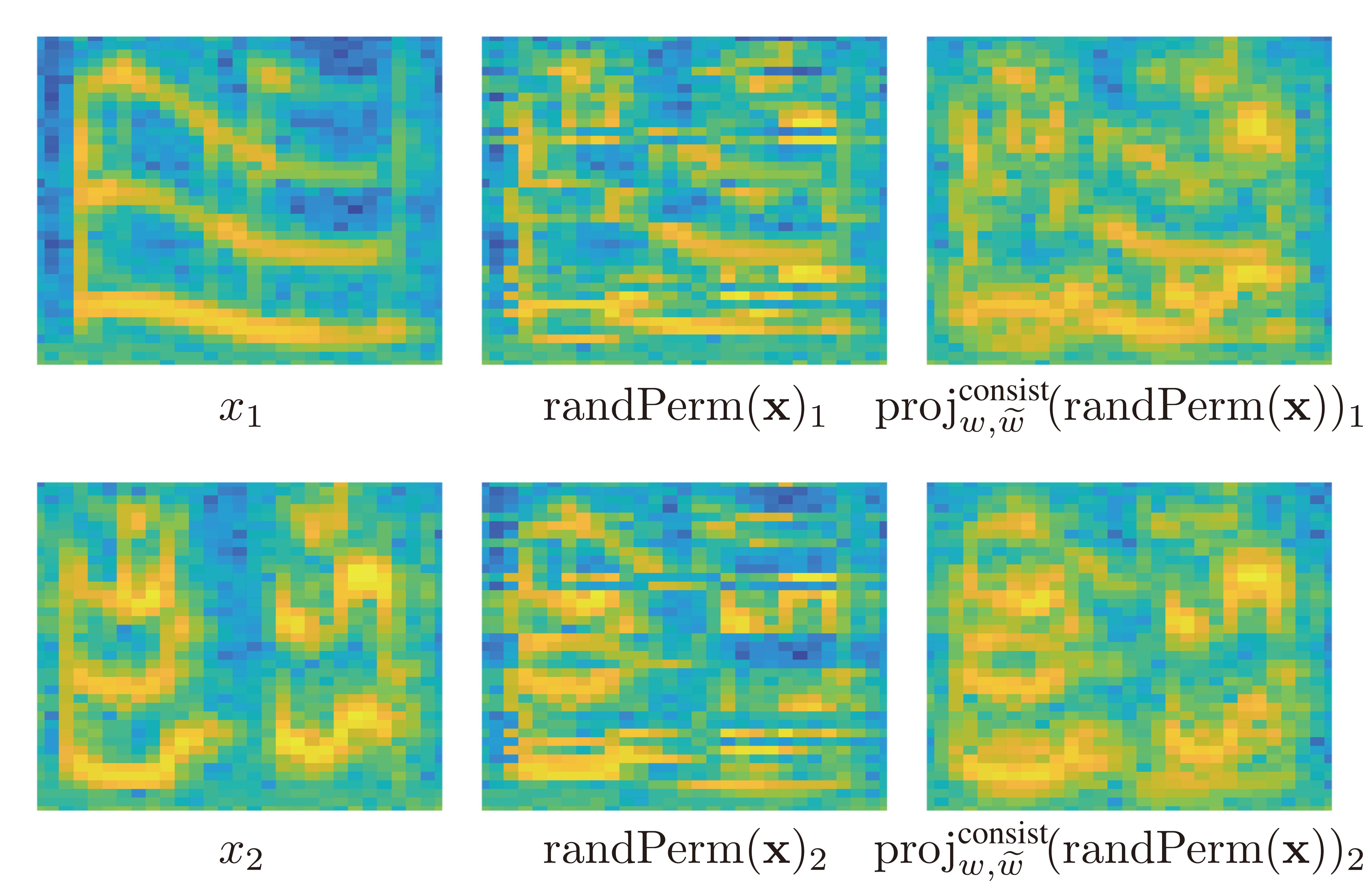}
    \vspace{-6pt}
    \caption{Examples of the relation between the spectrogram consistency and permutation problem. Clean speech signals (left) were randomly permuted (middle) and projected onto the consistent subspace (right).}
    \label{fig:speechPermutationExamples}
\end{figure}

The smoothing process of the inverse STFT clarifies why misalignment of the frequency-wise permutation in BSS is the problem.
For demonstrating that, perfectly separated speech signals are shown in Fig.\:\ref{fig:speechPermutationExamples}.
Original (consistent) spectrograms are illustrated in the left column, while their randomly permuted (inconsistent) versions are shown in the middle.
For each frequency, signals in both left and middle columns are perfectly separated because every time-frequency bin consists of only one of the speech signals.
However, the inverse STFT mixes up the signals in the time domain, which is illustrated in the right column.
It removes the non-smoothness caused by the permutation misalignment and spreads the signal components vertically as in Fig.\:\ref{fig:consistentExamples} because a consistent spectrogram must be smooth along frequency.
In order to separate the source signals, a BSS algorithm should be performed with consideration of such smoothing effect of the inverse STFT.

\section{Consistent Determined BSS}

In this section, a new formulation of determined BSS is proposed to take advantage of the spectrogram consistency.

\subsection{Independence-based Determined BSS}
\label{subsec:indepProb}

To estimate the $M\!\times\!M$ demixing matrices $\{\mathbf{W}[f]\}_{f=1}^F$ in Eq.~\eqref{eq:demixing}, the independence-based models are often considered.
By assuming statistical independence between the source signals, many of the determined BSS methods have been formulated as a minimization problem of the following form:
\begin{equation}
    \underset{\{\mathbf{W}[f]\}_{f=1}^F}{\text{Minimize}}\;\;\;\mathcal{P}(\mathbf{W}[f]\mathbf{x}[t,f])\,-\sum_{f=1}^F\log|\!\det(\mathbf{W}[f])|,
    \label{eq:logDetProblem}
\end{equation}
where $\mathcal{P}$ is a real-valued penalty function corresponding to the source model \cite{YatabeICASSP2018}, i.e., a function taking higher values for mixtures and lower values for correctly separated signals.
When this penalty function is separable for $f$ as
\begin{equation}
    \underset{\{\mathbf{W}[f]\}_{f=1}^F}{\text{Minimize}}\;\;\sum_{f=1}^F\mathcal{P}_{_{\!}f}(\mathbf{W}[f]\mathbf{x}[t,f])\,-\sum_{f=1}^F\log|\!\det(\mathbf{W}[f])|,
    \label{eq:sepLogDetProblem}
\end{equation}
it can be solved for each $f$ independently.
This separable formulation known as the frequency-domain ICA \cite{Comon1994,Smaragdis1998,hyvarinen2000,Buchner2005,Ono2010auxFDICA} is easier to solve but comes with a price of the permutation problem because the penalty function $\mathcal{P}_{_{\!}f}$ cannot distinguish the permutation which is an inter-frequency relation.
The usual technique for resolving the permutation is to design a non-separable function $\mathcal{P}$ to model inter-frequency relation of the source signals, e.g., IVA \cite{Hiroe2006,Kim2007,Ono2011auxIVA} and ILRMA \cite{Kitamura2015,Kitamura2016,KitamuraEURASIPJ2018} (see \cite{YatabeICASSP2018} for some explicit forms of $\mathcal{P}$).
That is, the key to solve the problem is to make the function $\mathcal{P}$ sensitive to misaligned permutation so that it can be detected and penalized.

\subsection{Vectorized Form of BSS Problem in Eq.~\eqref{eq:logDetProblem} \cite{YatabeICASSP2018}}
\label{subsec:indepProb}

For notational convenience, all demixing filters $\{\mathbf{W}[f]\}_{f=1}^F$ are represented by a single vector $\mathbf{w}$ as follows:
\begin{equation}
    \mathbf{w} = [\mathbf{w}[1]^{\mathrm{T}},\ldots,\mathbf{w}[F]^{\mathrm{T}}]^{\mathrm{T}},
    \quad
    \bigl(\,\mathbf{w}[f] = \mathcal{V}(\mathbf{W}[f])\,\bigr),
\end{equation}
where $\mathcal{V}$ is the operator converting a matrix into a vector,
\begin{equation}
    \mathcal{V}(\mathbf{W}[f]) \!=\! [W_{\!1,1}[f],\ldots, W_{\!1,M}[f],W_{\!2,1}[f],\ldots, W_{\!M,M}[f]]^{\mathrm{T}}\!\!\!.\!
\end{equation}
By defining a matrix $\mathbf{X}$ corresponding to the observation $\mathbf{x}$ as
\begin{align}
    \mathbf{X} &= \blkdiag(\boldsymbol{\chi}[1],\boldsymbol{\chi}[2],\ldots,\boldsymbol{\chi}[F]), \\[2pt]
    \boldsymbol{\chi}[f] &= \blkdiag(\chi[f], \chi[f], \ldots,\chi[f]), \;\; (\text{$M$ times}) \\[2pt]
    \chi[f] &= [\tau_1[f],\tau_2[f],\ldots,\tau_M[f]], \\
    \tau_m[f] &= [x_m[1,f],x_m[2,f],\ldots,x_m[T,f]]^{\mathrm{T}},
\end{align}
the BSS problem in Eq.~\eqref{eq:logDetProblem} can be compactly represented as
\begin{equation}
    \underset{\mathbf{w}}{\text{Minimize}}\;\;\;\mathcal{I}(\mathbf{w}) + \mathcal{P}(\mathbf{X}\mathbf{w}),
    \label{eq:PDS-BSSproblem}
\end{equation}
where $\blkdiag$ is the operator constructing a block-diagonal matrix by diagonally concatenating the input matrices, $\tau_m[f]$ is $T\!\times\!1$, $\chi[f]$ is $T\!\times\!M$, $\boldsymbol{\chi}[f]$ is $MT\!\times\!M^2$, $\mathbf{X}$ is $FMT\!\times\!FM^2$,
\begin{equation}
    \mathcal{I}(\mathbf{w}) = -\sum_{f=1}^F\sum_{m=1}^M\log\sigma_m(\mathcal{M}(\mathbf{w})[f]),
    \label{eq:defI(w)}
\end{equation}
$\mathcal{M}$ is the operator converting the vector back into the matrix, and $\sigma_m(\mathbf{W})$ is the $m$th singular value of $\mathbf{W}$.

\subsection{Proposed Formulation Realizing Consistent BSS}
\label{sec:propFromula}

From Fig.\:\ref{fig:speechPermutationExamples}, it is evident that BSS should be performed within the consistent subspace of spectrograms to avoid the mixing caused by the inverse STFT.
However, the ordinary BSS model in Eq.~\eqref{eq:PDS-BSSproblem} cannot manage that effect because the penalty function $\mathcal{P}$ does not distinguish the consistent and inconsistent components of the spectrogram.
Such BSS method measures the degree of separation based on both of the components, which may not promote source separation because it is possible to reduce the penalty by increasing the amount of inconsistent component ($\nu$ in Eq.~\eqref{eq:removeNu}).
Therefore, a BSS model should be insensitive to the inconsistent part of the filtered spectrogram $\mathbf{X}\mathbf{w}$.

To make a BSS method only sensitive to the consistent component of the filtered spectrogram, Eq.~\eqref{eq:PDS-BSSproblem} is slightly modified to include the projection in Eq.~\eqref{eq:defProj} as
\begin{equation}
    \underset{\mathbf{w}}{\text{Minimize}}\;\;\;\mathcal{I}(\mathbf{w}) + \mathcal{P}(\mathrm{proj}^{\text{consist}\!}_{w,\widetilde{w}}(\mathbf{X}\mathbf{w})).
    \label{eq:consistentBSSproblem}
\end{equation}
By this modification, the inconsistent components are ignored by the penalty function because they are removed by the projection.
The inter-frequency dependency of STFT can be handled by any function $\mathcal{P}$ since the projection $\mathrm{proj}^{\text{consist}\!}_{w,\widetilde{w}}(\cdot)$ spreads the components of the inputted spectrogram along frequency%
\footnote{Note that the proposed formulation can only partly resolve the permutation problem because the effect of the projection is local in the time-frequency plane.
Since a reasonable window function has small sidelobe, the major effect of spreading the frequency components is limited within its mainlobe typically supported in a few bins.
Therefore, the proposed method may not resolve the global or block-wise permutation.}.
The proposed composite function $\mathcal{P}(\mathrm{proj}^{\text{consist}\!}_{w,\widetilde{w}}(\cdot))$ can detect and penalize permutation misalignment, even when $\mathcal{P}$ is separable for $f$, based on the general \textit{signal-independent} property of the spectrogram.
This is an approach different from the conventional methods modeling the structure of source signals by \textit{signal-dependent} penalty functions.
Since the projection can be combined with any model represented by Eq.~\eqref{eq:PDS-BSSproblem}, it has potential to improve existing BSS methods%
\footnote{Because a consistent spectrogram directly corresponds to its time-domain counterpart, the proposed formulation measures the degree of separation in terms of the time-domain signals even though the mixing/demixing model is in the time-frequency domain. Therefore, the separation results obtained through the proposed modification should be different from the original ones post-processed by a permutation solver.}.

\subsection{Example Algorithm for Proposed Consistent BSS}

While any algorithm can be applied to handle the proposed BSS model in Eq.~\eqref{eq:consistentBSSproblem}, this paper utilizes a PDS algorithm as an example because its derivation is straightforward.
Since the projection in Eq.~\eqref{eq:defProj} is a bounded linear operator which can be written as a matrix, its composition with the matrix $\mathbf{X}$ can also be regarded as a matrix obtained by their matrix-matrix multiplication.
Therefore, a PDS algorithm for Eq.~\eqref{eq:consistentBSSproblem} can be obtained by simply replacing the matrix multiplication $\mathbf{X}\cdot$ of the algorithm in \cite{YatabeICASSP2018} by $\mathrm{proj}^{\text{consist}\!}_{w,\widetilde{w}}(\mathbf{X}\,\cdot\:)$ as shown in Algorithm~\ref{alg:PDS-consistentBSS-masking}, where the proximity operator is defined as
\begin{equation}
    \prox_{\mu g}[\,\mathbf{z}\,] = \arg\min_{\boldsymbol{\gamma}}\Bigl[\,g(\boldsymbol{\gamma}) + \frac{1}{2\mu}\left\|\mathbf{z}-\boldsymbol{\gamma}\right\|_2^2\,\Bigr].
    \label{eq:defProx}
\end{equation}

\begin{algorithm}[t]
\caption{Consistent determined BSS}
\label{alg:PDS-consistentBSS-masking}
\begin{algorithmic}[1]
\STATE \textbf{Input:} $\mathbf{X}$, $\mathbf{w}^{[1]}$, $\mathbf{y}^{[1]}$, $\mu_1$, $\mu_2$, $\alpha$
\STATE \textbf{Output:} $\mathbf{w}^{[K+1]}$
\FOR{$k = 1, \ldots, K$}
\STATE $\widetilde{\mathbf{w}} = \prox_{\mu_1 \mathcal{I}}[\;\mathbf{w}^{[k]}-\mu_1\mu_2 \mathbf{X}^{\mathrm{H}}\mathrm{proj}^{\text{consist}\!}_{\widetilde{w},w}(\mathbf{y}^{[k]})\;]$
\vspace{2pt}
\STATE $\,\mathbf{z}\, = \mathbf{y}^{[k]} + \mathrm{proj}^{\text{consist}\!}_{w,\widetilde{w}}(\mathbf{X}(2\widetilde{\mathbf{w}}-\mathbf{w}^{[k]}))$
\vspace{2pt}
\STATE $\,\widetilde{\mathbf{y}} = \,\mathbf{z} - \prox_{\frac{1}{\mu_2}\mathcal{P}}[\,\mathbf{z}\,]$
\STATE $\,\mathbf{y}^{[k+1]} = \alpha\widetilde{\mathbf{y}}+(1-\alpha)\mathbf{y}^{[k]}$
\vspace{1pt}
\STATE $\mathbf{w}^{[k+1]} = \alpha\widetilde{\mathbf{w}}+(1-\alpha)\mathbf{w}^{[k]}$
\ENDFOR
\end{algorithmic}
\end{algorithm}

This algorithm can be applied to many BSS models by only changing $\prox_{\mathcal{P}/\mu_2\!}$ in the 6th line (see \cite{YatabeICASSP2018} for details).
Thus, a conventional method can be easily extended to its consistent version if the corresponding proximity operator is available.
It can also be extended to a general time-frequency mask \cite{YatabeICASSP2019}, which should be more convenient than the other algorithms for testing source models thanks to the easiness of the code modification.
Note that this algorithm is merely an example, and it is possible to design a faster algorithm for a specific model (defined by fixing $\mathcal{P}$), which is left as a future work.

\section{Experiment}

For the experiment, as well-understood examples of the BSS models, the frequency-domain ICA based on the Laplace distribution and IVA based on the spherical Laplace distribution were considered.
In the case of the Laplace ICA, the penalty function $\mathcal{P}$ is the $\ell_1$-norm $\|\cdot\|_1$, and its proximity operator is given by the well-known soft-thresholding operator:
\begin{equation}
    \bigl(\prox_{\lambda\|\cdot\|_1}[\,\mathbf{z}\,]\bigr)_m[t,f] = \left(1-\lambda/|z_m[t,f]|\right)_+z_m[t,f],
    \label{eq:proxL1}
\end{equation}
where $(\cdot)_+=\max\{\cdot,0\}$.
Similarly, for the Laplace IVA, the penalty function $\mathcal{P}$ is the $\ell_{2,1}$-mixed norm $\|\cdot\|_{2,1}$, and its proximity operator is the following group thresholding:
\begin{equation}
    \bigl(\prox_{\lambda\|\cdot\|_{2,1}}[\,\mathbf{z}\,]\bigr)_m[t,f] = \left(1-\lambda/\zeta_m[t]\right)_+z_m[t,f],
    \label{eq:proxL21}
\end{equation}
where $\zeta_m[t]=(\sum_{f=1}^F|z_m[t,f]|^2)^{\frac{1}{2}}$.
By inserting these operators into the 6th line of Algorithm~\ref{alg:PDS-consistentBSS-masking}, the consistent versions of the ICA and IVA algorithms are obtained.
Their ordinary versions are also obtained by ignoring the projection in the 4th and 5th lines as shown in \cite{YatabeICASSP2018}.

The proposed method was tested by applying it to speech mixtures as in \cite{Kitamura2016}.
The database used in this experiment was a part of SiSEC (\texttt{UND} task) \cite{SAraki2012_SiSEC}.
The ICA and IVA with and without the proposed projection were evaluated for the two cases: 2-channel and 3-channel separation.
For the 2-channel signals, 12 speech mixtures (\texttt{liverec}) contained in \texttt{dev1} and \texttt{dev2}, which include female/male speech with the reverberation time 130\,ms/250\,ms and the microphone spacing 1\,m/5\,cm, were utilized.
The first two speech sources for each mixture were chosen to make the task determined ($N\!=\!M\!=\!2$) as done in \cite{Kitamura2016}.
For the 3-channel signals, 8 speech mixtures in \texttt{dev3}, which include female/male speech with the reverberation time 130\,ms/380\,ms and the microphone spacing 50\,cm/5\,cm, were utilized.
The first three speech sources for each mixture were chosen to make the task determined ($N\!=\!M\!=\!3$).
See \cite{SAraki2012_SiSEC} for the other conditions.
The tightened half-overlapping 1024-points-long Hann window was used for STFT.
The parameters in Algorithm~\ref{alg:PDS-consistentBSS-masking} were set to $\mu_1\!=1$, $\mu_2\!=1$, $\alpha=1.75$, and $K=2000$.
The initial value of the demixing matrices $\mathbf{w}^{[1]}$ was set to the identity matrices ($\mathbf{W}[f]=\mathbf{I}$ for all $f$), and that of $\mathbf{y}$ was the zero vector.

\begin{figure}[!t]
    \vspace{-8pt}
    \centering
    \includegraphics[width=0.98\columnwidth]{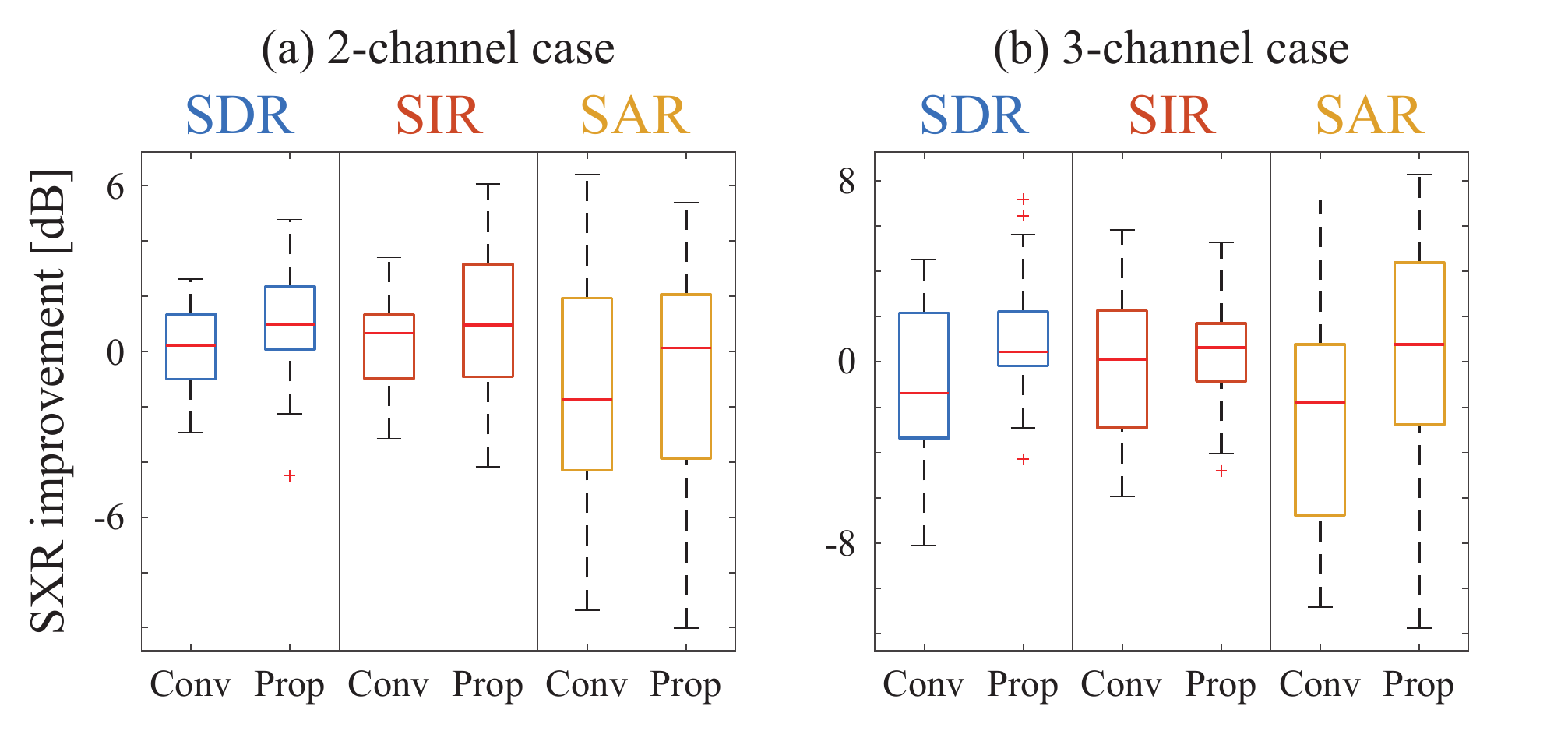}
    \vspace{-10pt}
    \caption{Box plot of SDR/SIR/SAR improvement of the Laplace ICA ($\lambda=0.1$) with (right) and without (left) the proposed projection. The central lines are the median, and the edges of the box represent the 25th and 75th percentiles.}
    \label{fig:boxL1}
\end{figure}

\begin{figure}[!t]
    \centering
    \includegraphics[width=0.98\columnwidth]{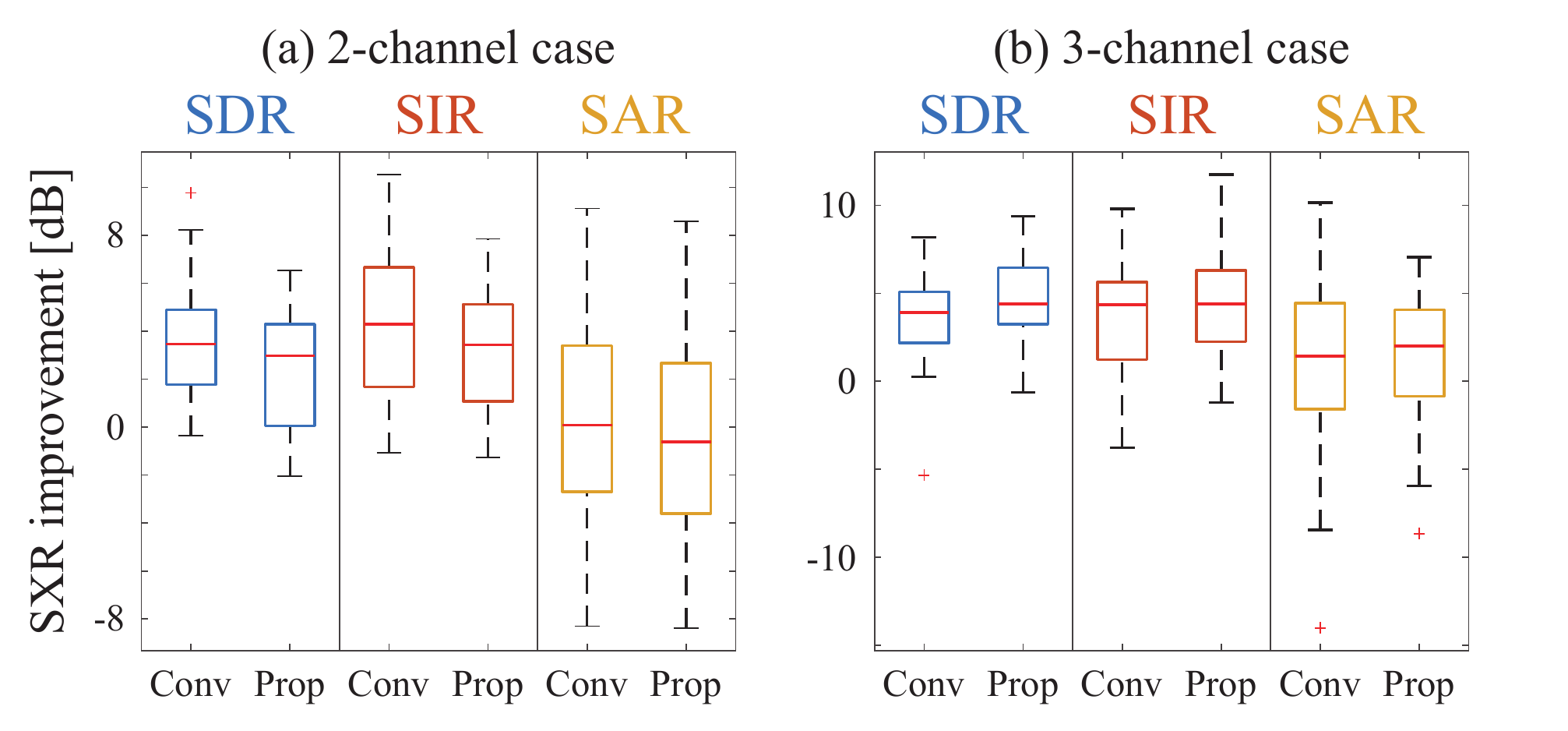}
    \vspace{-10pt}
    \caption{Box plot of SDR/SIR/SAR improvement of the Laplace IVA ($\lambda=1$) with (right) and without (left) the proposed projection.}
    \label{fig:boxL21}
\end{figure}

The experimental results for the ICA are summarized in Fig.\:\ref{fig:boxL1} (the conventional and proposed methods are placed next to each other for comparison).
In this experiment, no permutation solver was utilized, and therefore the separation must suffer from the permutation problem.
Such situation can be read from Eq.~\eqref{eq:proxL1} because it is a frequency-independent thresholder and cannot do anything to the inter-frequency misalignment.
From the figure, it can be seen that the proposed projection improved the scores for all cases.
This result indicates that it is possible to partly solve the permutation problem by only considering the spectrogram consistency within the BSS algorithm as expected in Section~\ref{sec:propFromula}.

The experimental results for the IVA are summarized in Fig.\:\ref{fig:boxL21}.
This experiment investigated the influence of the projection on the time-related separation cues because IVA is only sensitive to the time-directional fluctuation of the signal, which can be seen from Eq.~\eqref{eq:proxL21} that squeezes all frequency-dependent information.
As in the figure, while the scores for the 2-channel case were worsened, those for the 3-channel case were improved.
The projection spreads the energy in both frequency and time directions as in Fig.\:\ref{fig:consistentExamples}.
The results indicate that the projection acts favorably on the frequency-related issue although further investigation is required to reveal its effect on the time-related information.

\vspace{-2pt}
\section{Conclusions}

In this paper, a new formulation of determined BSS was proposed as a first step to incorporate the general property of STFT called spectrogram consistency.
By proposing an algorithm for handling that, the potential of spectrogram consistency for improving BSS was experimentally shown.
The approach to the fusion of spectrogram consistency and BSS is not limited to the method proposed in this paper, and the other methods in phase-aware signal processing and phase reconstruction should be able to improve the performance of BSS.
Future works include investigating a more sophisticated model and method as well as developing a faster algorithm.



\end{document}